\documentclass{article}
\usepackage{spconf}
\usepackage{hyperref}
\usepackage{url}
\usepackage{graphicx}
\usepackage{epstopdf}
\usepackage{algorithm}
\usepackage{algorithmicx}
\usepackage{algpseudocode}
\usepackage{cite}
\usepackage{color}
\usepackage{amsmath,amssymb,amsfonts,amsthm}
\usepackage{bm}
\usepackage{setspace}
\usepackage{multirow}
\usepackage{mathtools}
\usepackage{wrapfig}
\usepackage{subcaption}

\newcommand{\Lb}{\left[}
\newcommand{\Rb}{\right]}
\newcommand{\lb}{\left(}
\newcommand{\rb}{\right)}

\newcommand{\bA}{\mathbf{A}}
\newcommand{\bV}{\mathbf{V}}
\newcommand{\bW}{\mathbf{W}}

\newcommand{\bD}{\mathbf{D}}

\newcommand{\bC}{\mathbf{C}}

\newcommand{\bone}{\mathbf{1}}

\newcommand{\by}{\mathbf{y}}

\newcommand{\ba}{\mathbf{a}}
\newcommand{\bw}{\mathbf{w}}

\newcommand{\bZ}{\mathbf{Z}}

\newcommand{\bX}{\mathbf{X}}
\newcommand{\bY}{\mathbf{Y}}
\newcommand{\bI}{\mathbf{I}}

\newcommand{\bQ}{\mathbf{Q}}

\newcommand{\trace}{\textnormal{trace}}

\newcommand{\cV}{\mathcal{V}}
\newcommand{\cE}{\mathcal{E}}
\newcommand{\row}{\textnormal{row}}

\begin{document}
\title{Multi-centrality Graph Spectral Decompositions and their Application to Cyber Intrusion Detection}

\name{Pin-Yu Chen$^{\star}$ \qquad Sutanay Choudhury$^{\dagger}$
	\thanks{This work was partially supported by the Consortium for Verification Technology under Department of Energy National Nuclear Security Administration award number DE-NA0002534 and by the 
		Asymmetric Resilient Cyber Security initiative at Pacific Northwest National Laboratory, which is operated by Battelle Memorial Institute.}
	 \qquad Alfred O. Hero III$^{\star}$,~Fellow,~IEEE	
	}
\address{$^{\star}$Department of Electrical Engineering and Computer Science, University of Michigan, Ann Arbor, USA \\ \{pinyu, hero\}@umich.edu \\
$^{\dagger}$Pacific Northwest National Laboratory \\ Sutanay.Choudhury@pnnl.gov}



\ninept 
\maketitle
\begin{abstract}
	Many modern datasets can be represented as graphs and hence spectral decompositions such as graph principal component analysis (PCA) can be useful. Distinct from previous graph decomposition approaches based on subspace projection of a single topological feature, e.g., the Fiedler vector of centered graph adjacency matrix (graph Laplacian), we propose spectral decomposition approaches to graph PCA and graph dictionary learning that integrate multiple features, including graph walk statistics, centrality measures and graph distances to reference nodes. In this paper we propose a new PCA method for single graph analysis, called multi-centrality graph PCA (MC-GPCA), and a new dictionary learning method for ensembles of graphs, called multi-centrality graph dictionary learning (MC-GDL), both based on spectral decomposition of multi-centrality matrices. As an application to cyber intrusion detection, MC-GPCA can be an effective indicator of anomalous connectivity pattern and MC-GDL can provide discriminative basis for attack classification.
\end{abstract}

\section{Introduction}
Many real-world data ranging from physical systems, social interactions, network flows, knowledge graphs to biological and chemical reactions are often represented as graphs, especially for anomaly or community detection \cite{Newman10NetworkIntro,noble2003graph,hogan2013towards,CPY14ComMag,Joslyn2013,CPY14spectral,Akoglu15Graphanomalydetection,Miller15AnomalousSubgraph,Oler2015} and graph signal processing \cite{Shuman13,Bertrand13Mag,anis2014towards,Wang15localset,SihengChen15signalrecovery}.
Dimensionality reduction methods on graphs allow one to decompose a graph into principal components using a spectral decomposition of the graph adjacency or graph Laplacian matrix. In this paper we propose a general framework to dimensionality reduction based on spectral decomposition of a matrix composed of many different graph centrality statistics. This general framework leads to a single-graph decomposition method that extends graph principal components analysis (PCA) and a graph-ensemble decomposition method  that extends dictionary learning. These methods are applicable to both directed and undirected graphs with edge weights and are based on a spectral decomposition, specifically the singular value decomposition (SVD), of a matrix composed of multiple graph centrality statistics. The proposed methods are denoted multi-centrality graph PCA (MC-GPCA) and multi-centrality graph dictionary learning (MC-GDL), respectively. By integrating multiple descriptions of graph centrality,  the proposed methods provide graph community detection and graph structure learning that are significantly more robust to noise and variation affecting graph connectivity structures.

 In \cite{saerens2004principal}, a kind of graph PCA is performed on the distance matrix of average commute time between nodes. In \cite{jiang2013graph}, PCA can also be performed on the graph Laplacian matrix of nodal similarities. In \cite{ShahidKBBV15}, the graph Laplacian matrix is used as a smooth regularization function for robust PCA on graphs. 
In \cite{lakhina2004diagnosing,terrell2005multivariate}, PCA is performed on the matrix of origin-destination traffics. In \cite{Thanou14TSP,Zhang12graphdictionary,Thanou15ICASSP}, dictionary learning methods for   graph signals are proposed based on the graph Laplacian matrix.
Dictionary learning, also known as sparse coding, linear unmixing and matrix factorization, has been applied to collections of images, audio, and graph signals to learn low dimensional representations that give a sparse approximation to the entire collection. Dictionary learning finds a low rank factored-matrix approximation to the observation matrix, whose columns span this collection.   Many different methods for this approximation problem have been proposed
\cite{lee2006efficient,jenatton2010proximal}. Among the simplest methods is the K-SVD approach \cite{aharon2006KSVD} which uses a spectral decomposition to determine the best low rank approximation to the observed matrix. For the purposes of illustration, in this paper we adopt this latter spectral approach for learning a dictionary spanning an ensemble of graphs.

More often graph PCA and graph dictionary learning approaches start with a set of raw multivariate data samples, create a similarity (or dissimilarity) graph of the data samples, and aim to learn a low-dimensional or sparse representation of the original multivariate dataset.
When applied to graph data, these methods are often limited to graphs that are weighted, undirected and connected, which may not be feasible for applications such as cyber network data analysis. Furthermore, these methods often accomplish graph decomposition based on a single measure of centrality, e.g.,  betweeness centrality \cite{Freeman77}, closeness centrality \cite{Sabidussi66Closeness}, ego centrality \cite{Everett05Ego}, or eigenvector centrality \cite{Newman10NetworkIntro}.    
In this paper we introduce graph spectral decomposition methods that combine multiple centrality measures such as graph walk statistics and graph distances as structural features 
and apply them to different graph types including weighted, directed and disconnected graphs. 
The proposed MC-GPCA method decomposes a single graph utilizing multiple centrality features, achieving dimensionality reduction and feature decorrelation  of the graph.
The proposed MC-GDL performs dictionary learning across a population of graphs using multiple centrality features to learn the atoms of the dictionary   and the corresponding coefficients to represent each individual graph in terms of its projection onto the dictionary.
 Applying our approach to cyber intrusion detection, we use MC-GPCA to define a structural difference score (SDS) that reflects structural variations within a graph and we use 
 MC-GDL to learn discriminative structural atoms for classifying the presence of cyber attacks.

\section{Structural Feature Extraction on Graphs}
\label{sec_structural_feature}
Here we describe three categories of generic structural features that can be extracted from a graph, namely graph walk statistics, centrality measures and internode distances. The utility of the introduced features with respect to different graph types, including weighted, directed and disconnected graphs, is summarized in Table \ref{table_feasiblity_structural feature}.
While not investigated in this paper, application-specific features such as website hit rates, social interaction frequency, source-destination traffics can also be leveraged as structural features. 
Without loss of generality a graph $G=(\cV,\cE)$ can be characterized by two $n \times n$ matrices $\bA$ and $\bW$ representing the adjacency and weight matrix, respectively, where $\cV$ ($\cE$) is the set of nodes (edges), and $n$ is the total number of nodes (i.e., graph size). $\bA$ is a binary matrix such that its entry $[\bA]_{ij}=1$ if there is an edge connecting from node $i$ to node $j$, and $[\bA]_{ij}=0$ otherwise. Throughout this paper we consider graphs with nonnegative edge weights such that  $\bW$ is a nonnegative matrix, where its entry $[\bW]_{ij} \geq 0$ if $[\bA]_{ij}=1$,  and  
$[\bW]_{ij} = 0$ otherwise.

\subsection{Graph walk statistics}
Graph walk statistics include commute time and cover time\cite{lovasz1993random}, graph diffusion \cite{gomez2010inferring}, hitting times \cite{galluccio2013clustering}, and hop walks. In this paper we focus on hop walk statistics. An $h$-hop walk of a node on a graph is a path starting from the node and traversing through (possibly repeated) $h$ edges. An $h$-hop walk weight is defined as the sum of edge weights of the corresponding path.
We consider the number and total weight of $h$-hop  walks of each node as  features since they entail the structural information of nodal reachability relative to its $h$-hop vicinity.
In principle one should extract graph walk statistics from $h=1$ to at least $h=\textit{graph~diameter}$ hops as structural features, where graph diameter is the largest shortest path hop count between any node pairs in all connected components of a graph.
We propose an efficient iterative computation method to incrementally computes these two structural features with respect to the hop count number $h$:~\\
	$\bullet$ \textbf{Iterative computation of number of $h$-hop walks}~\\ 
	Let $\bA^h$ denote the matrix product of $h$ copies of $\bA$. Observe that the entry of $\bA^2$, 
	$[\bA^2]_{ij}=\sum_k [\bA]_{ik} [\bA]_{kj}$, is the number of $2$-hop walks from $i$ to $j$. Extending this result to $\bA^h$ we have 
	$[\bA^h]_{ij}$ being the number of $h$-hop walks from $i$ to $j$. Let $\ba^{(h)}=\bA^h \bone_n$ be a column vector where its entry $[\ba^{(h)}]_i$ is the number of $h$-hop walks starting from $i$ and $\bone_n$ denotes the $n \times 1$ column vector of ones. Then $\ba^{(h+1)}$ can be computed by the matrix-vector product iteration
	\begin{align}
	\ba^{(h+1)}=\bA^{h+1} \bone_n = \bA \cdot \bA^{h} \bone_n =  \bA \ba^{(h)}.
	\end{align}		
	$\bullet$ \textbf{Iterative computation of total $h$-hop walk weight}~\\
	Let $\bW^{(h)}$ be an $n \times n$ matrix such that its entry $[\bW^{(h)}]_{ij}$ is the sum of all $h$-hop walk weights from node $i$ to node $j$.
	Then we have
	\begin{align}
	[\bW^{(h+1)}]_{ij}&= \sum_{k \in \cV}  \lb [\bW]_{ik} \cdot [\bA^{h}]_{kj} + [\bW^{(h)}]_{kj} \rb \cdot \bA_{ik}  \nonumber \\
	&=\sum_{k \in \cV}    [\bW]_{ik}   \cdot [\bA^{h}]_{kj}  + \sum_{k \in \cV}   [\bW^{(h)}]_{kj} \cdot \bA_{ik}  \nonumber \\
	&=[\bW \bA^h + \bA \bW^{(h)}]_{ij},
	\end{align}
	where we use $[\bW]_{ik} \cdot [\bA]_{ik}=[\bW]_{ik}$.
	Let $\bw^{(h)}=\bW^{(h)} \bone_n$ denote a column vector such that its entry $[\bw^{(h)}]_i$ is the total $h$-hop walk weight starting from node $i$. Then $\bw^{(h+1)}$  can be computed by
		\begin{align}
		\bw^{(h+1)}=\Lb \bW \bA^{h} + \bA \bW^{(k)} \Rb \bone_n 
		=\bW \ba^{(h)}+ \bA \bw^{(h)}.
		\end{align}

\begin{table}[]
	\centering
	\caption{Utility of the introduced structural features.}
	\label{table_feasiblity_structural feature}
	\begin{tabular}{l|ccc}
		\hline
		Feature / Graph Type                                                          & \multicolumn{1}{l}{Weighted} & \multicolumn{1}{l}{Directed} & \multicolumn{1}{l}{Disconnected} \\ \hline
		\# of $h$-hop graph walks                                     & \checkmark                   & \checkmark                   & \checkmark                       \\ \hline
		total $h$-hop walk weight                                         & \checkmark                   & \checkmark                   & \checkmark                       \\ \hline
		degree                                                            & \checkmark                   & \checkmark                   &   \checkmark                                    \\ \hline
		betweenness                                                       & \checkmark                   & \checkmark                   &                                  \\ \hline
		closeness                                                         & \checkmark                   & \checkmark                   &                                  \\ \hline
		eigenvector centrality                                         & \checkmark                   & \checkmark                   & \checkmark                       \\ \hline
		ego                                                               & \checkmark                   & \checkmark                   &         \checkmark                   \\ \hline
		LFVC                                                           & \checkmark                   &                              & \checkmark                       \\ \hline
		graph distance                                & \checkmark                   & \checkmark                   &                                  \\ \hline
	\end{tabular}
\end{table}

\subsection{Centrality measures }
\label{subsec_centrality_measure}
A centrality measure is a quantity that evaluates the level of importance or influence of a node in a graph and it reflects certain topological characteristics. 
Here we introduce several centrality measures, which will be used in the sequel to define feature sets associated with a graph or a set of graphs.
~\\	
	$\bullet$  \textbf{Degree.} Degree is defined as the number of edges associated with a node. It can be extended to directed graphs by considering the number of edges connecting to (from) a node as in-degree (out-degree).	
	~\\		
	$\bullet$ \textbf{Betweenness \cite{Freeman77}.} Betweenness is the fraction of shortest paths passing through a node relative to the total number of shortest paths in the graph. It is infeasible for disconnected graphs since it is based on shortest path distance.
	The betweenness of node $i$
	is defined as
	\begin{align}
	 \text{betweenness}(i)=\sum_{k \in \cV,k \neq i} \sum_{j \in \cV, j \neq i, j > k} \frac{\sigma_{kj}(i)}{\sigma_{kj}},
	\end{align}
	 where $\sigma_{kj}$ is the total number of shortest paths from $k$ to $j$ and $\sigma_{kj}(i)$ is the number of such shortest paths passing through $i$.
	 ~\\	
	$\bullet$  \textbf{Closeness \cite{Sabidussi66Closeness}.} Closeness is associated with the shortest path distances of a node to all other nodes.
	Let $\rho(i,j)$ denote the shortest path distance between node $i$ and node $j$ in a connected graph. Then 
	\begin{align}
	\text{closeness}(i)=\frac{1}{{\sum_{j \in \mathcal{V}, j\neq i} \rho(i,j)}}.
	\end{align}	
	~\\	
	$\bullet$  \textbf{Eigenvector centrality \cite{Newman10NetworkIntro}.}   Eigenvector centrality of node $i$ is the $i$-th entry of the eigenvector associated with the largest eigenvalue of the weight matrix $\bW$. It
	is defined as 
	\begin{align}
		\text{eigenvector centrality}(i)=\lambda_{\max}^{-1} \sum_{j\in \mathcal V} \mathbf [\bW]_{ji}[\boldsymbol{\xi}]_{j},
	\end{align}
 where $(\lambda_{\max}$,~$\boldsymbol{\xi})$   is the largest right eigenpair of $\bW^T$.
 ~\\
	$\bullet$  \textbf{Ego centrality \cite{Everett05Ego}.}  Ego centrality can be viewed as a local version of betweenness that computes the shortest paths between its neighboring nodes. 	Let $d_i$ denote the degree of node $i$, $\bW(i)$ denote the $(d_i+1)~\times~(d_i+1)$ local weight matrix of node $i$, $\bI$ be the identity matrix, and let $\circ$ denote entrywise matrix product. Ego centrality is defined as 
	\begin{align}
			\text{ego}(i)=\sum_{k \in \cV} \sum_{j \in \cV,j>k} \frac{1}{{\left[\bW^2(i) \circ \left(\bI-\bW(i)\right) \right]_{kj}}}.
	\end{align}
	$\bullet$  \textbf{Local Fiedler Vector Centrality (LFVC) \cite{CPY14ICASSP}.} LFVC is a centrality measure that evaluates the structural importance of a node regarding graph connectivity.
	Let $\by$ denote the eigenvector associated with the smallest nonzero eigenvalue of the graph Laplacian matrix. LFVC is defined as	
	\begin{align}
	 \text{LFVC}(i)=\sum_{j \in \mathcal{N}_i}([\by]_i-[\by]_j)^2,
	\end{align}
where $\mathcal{N}_i$ is the set of nodes connecting to or from $i$ (i.e., neighbors). 

\begin{figure*}[]
	\centering
	\begin{subfigure}[b]{0.33\textwidth}
		\includegraphics[width=\textwidth]{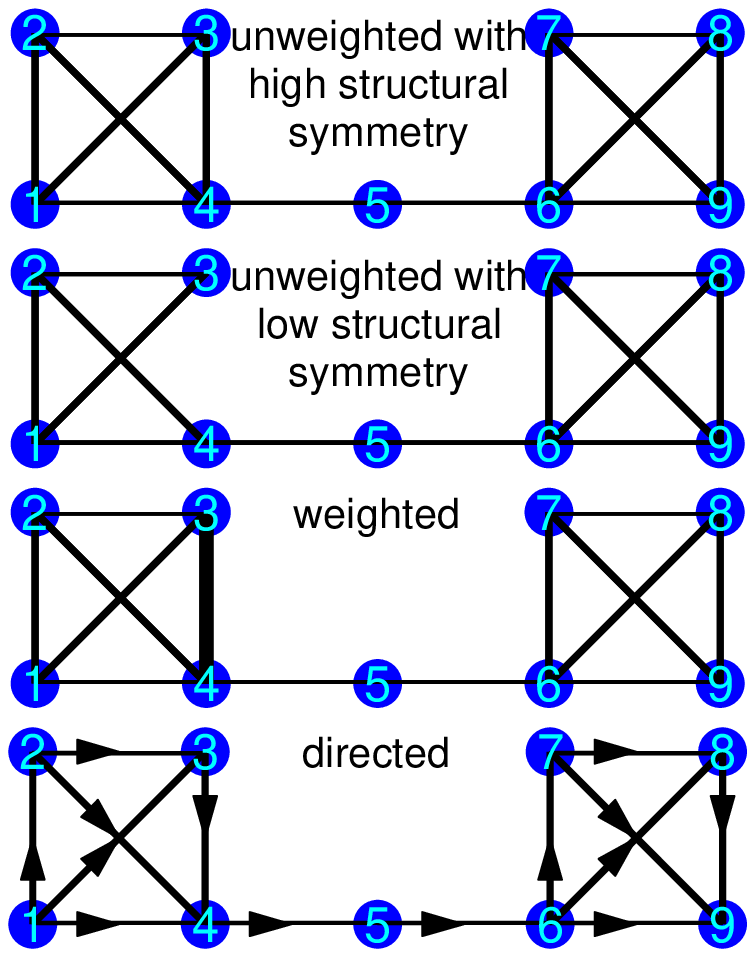}
		\caption{Four example graphs }
		\label{Fig_GPCA_demo}
	\end{subfigure}%
	\centering
	\begin{subfigure}[b]{0.33\textwidth}
		\includegraphics[width=\textwidth]{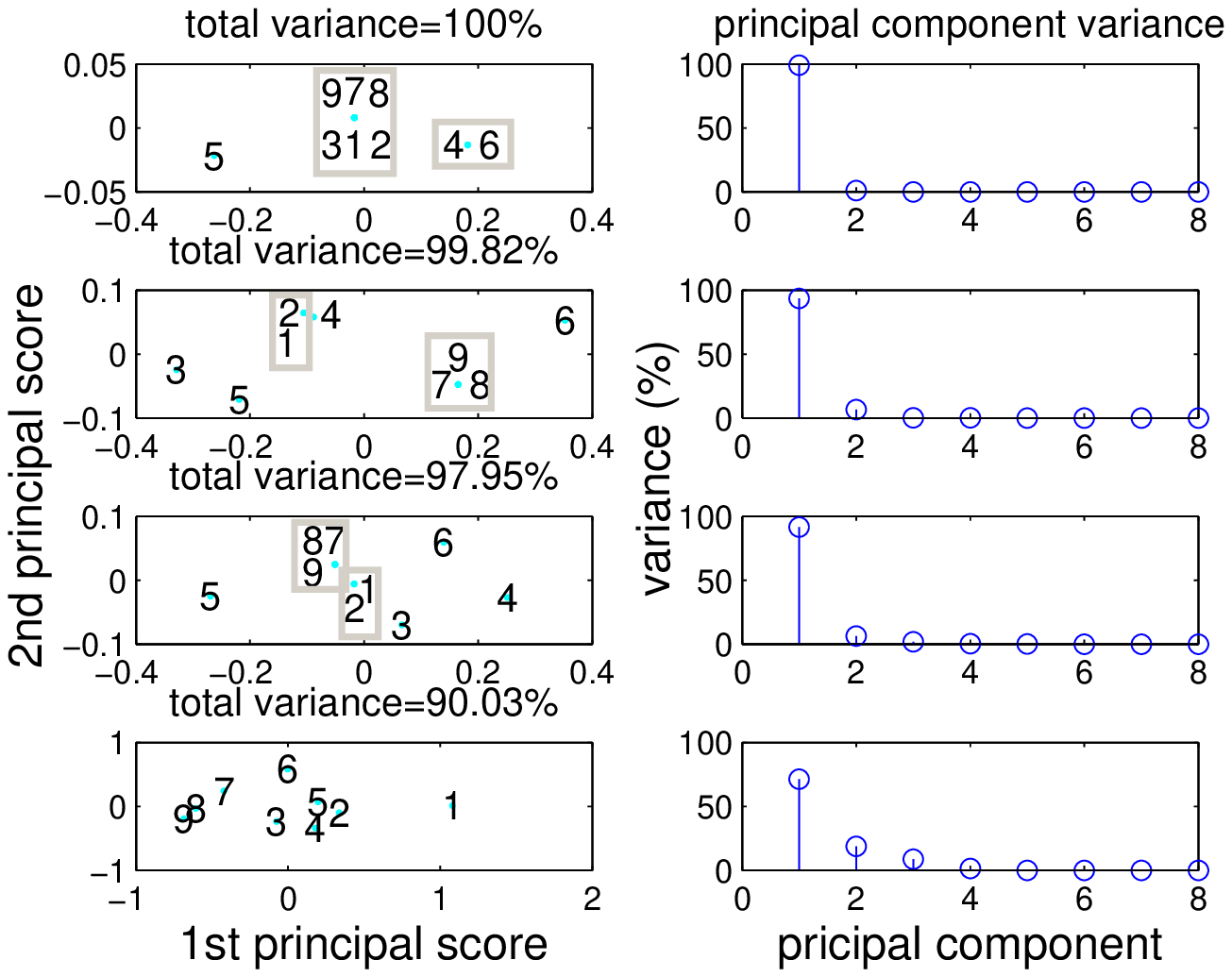}
		\caption{MC-GPCA without reference nodes}
		\label{Fig_GPCA_demo_no_ref}
	\end{subfigure}%
	\centering
	\begin{subfigure}[b]{0.33\textwidth}
		\includegraphics[width=\textwidth]{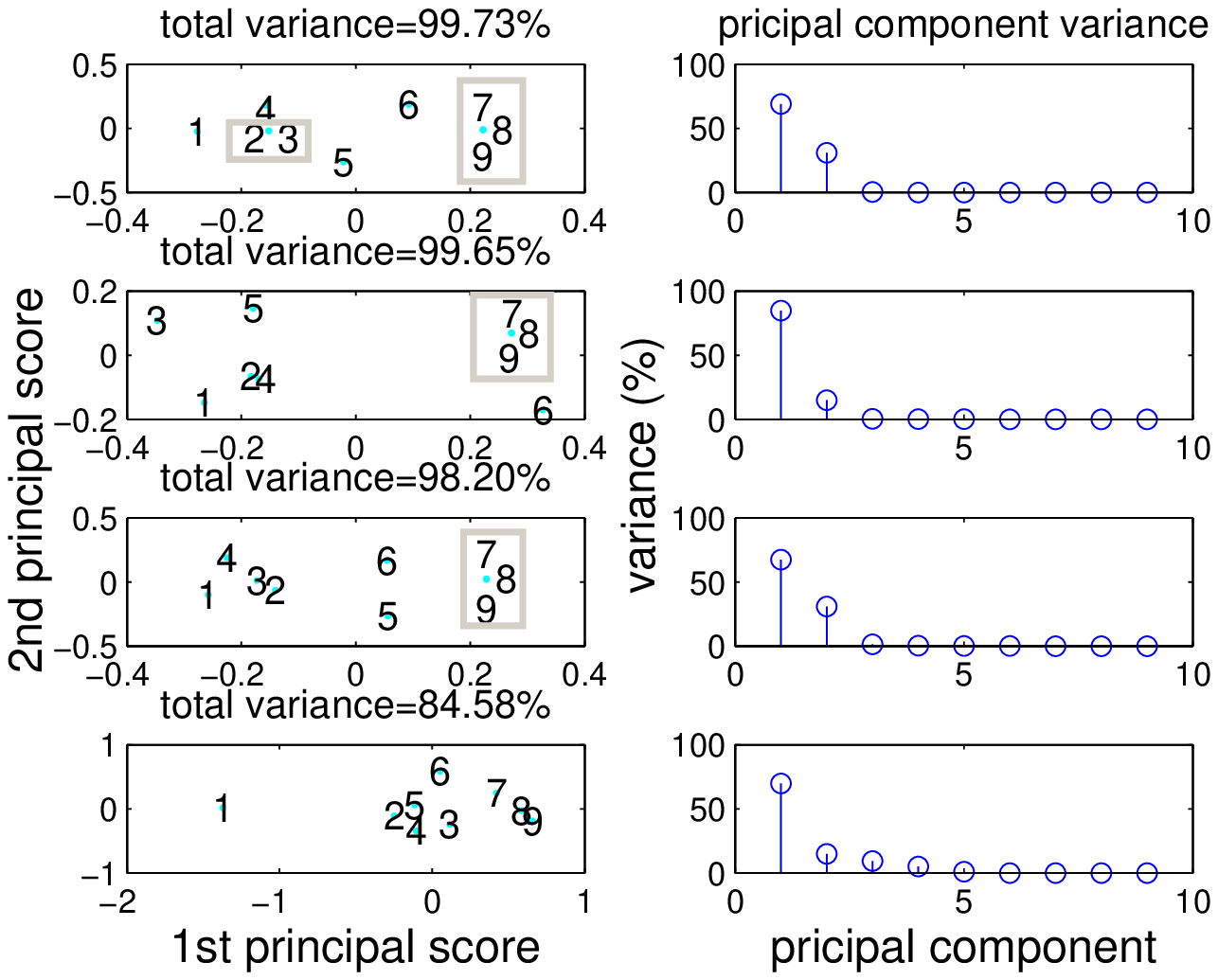}
		\caption{MC-GPCA with one reference node}
		\label{Fig_GPCA_demo_ref}
	\end{subfigure}%
	\caption{Illustration of sensitivity of proposed MC-GPCA algorithm to structural perturbations. Each node on a graph is represented by a 2-dimensional structural coordinate. Nodes marked by a gray box have identical MC-GPCA score.  
	}
	\label{Fig_SGPCA_demo}
\end{figure*}

\subsection{Graph distances to a set of reference nodes}
We propose to use graph distances of each node to a set of reference nodes as structural features that compensate the insufficiency of graph walk statistics and centrality measures when one performs MC-GPCA on graphs with high structural symmetry. For example, consider a star-like graph where the central node is a singleton and each leaf node is an identical clique (i.e., a complete graph). All edges in the graph are undirected and have identical weight. Therefore this graph has high structural symmetry and apparently the nodes of identical structural property (e.g., connected to the central node or not) have the same graph walk statistics and centrality measures. To resolve the ambiguity of graph walk statistics and centrality measures due to high structural symmetry in graphs we use the shortest path distance of each node  to the selected $r$ reference nodes as the $r$ additional structural features. In the example of the star-like graph with high structural symmetry, if $r=1$ then selecting any but the central node as a reference node can yield distinguishable structural features due to difference in shortest path distance to the reference node. 
The reference nodes are selected according to a user specified criterion, e.g. the nodes of maximal degrees.

\section{Methodology }
The extracted centrality features introduced in Sec. \ref{sec_structural_feature} can be represented as an $n \times p$ matrix $\bX$, where $n$ is the graph size, $p$ is the number of extracted features and each column of $\bX$ corresponds to a particular centrality feature that is normalized to have unit norm. The multi-centrality feature matrix $\bX$ is then centered by subtracting the row-wise empirical average from each row. 
\subsection{Multi-centrality graph PCA (MC-GPCA)}
In analogy to standard graph PCA, which is applied to the graph Laplacian matrix, MC-GPCA is PCA applied to $\bX$.  
PCA can be formulated as finding an orthonormal transformation $\bQ$ on  $\bX$ such that after transformation the multi-centrality feature matrix $\mathbf X$ is represented by an $n \times q~(q \leq p)$ matrix $\bY=\bX \bQ$ that maximally preserves the total data variance $\trace(\bY^T \bY)/n$,
 where $\trace( \cdot )$ denotes the sum of diagonal entries of a matrix and $\bQ$ is a $p \times q$ matrix such that $\bQ^T \bQ=\bI$. Such a matrix  $\bQ$ can be obtained by solving the $q$ right singular vectors associated with the $q$  largest singular values of $\bX$, which is denoted by a $p \times q$ matrix $\bV_q$. Moreover, the total variance of $\bY$ is equivalent to the sum of the $q$ squared largest singular values of $\bX$ divided by $n$. 
 Therefore using MC-GPCA we obtain $n$ $q$-dimensional coordinates representing structural scores with respect to the $q$ principal components (i.e., columns of $\bV_q$). The algorithm for MC-GPCA is summarized in \textbf{Algorithm} \ref{algo_strucutural_PCA}.

\subsection{Structural difference score (SDS)}
 We use these structural coordinates  (i.e., each row of $\bY=\bX \bV_q$) to define a structural difference score (SDS) for each node in a graph. The SDS of node $i$ is associated with the total squared Euclidean distance to its neighboring nodes $\mathcal{N}_i$ and its number of edges (i.e., degree $d_i$), which is defined as 
\begin{align}
\label{eqn_SDS}
\text{SDS}(i)=\frac{\sum_{j \in \mathcal{N}_i} \| \row_i(\bY) - \row_j(\bY)  \|^2}{d_i+1},
\end{align}
where $\row_i(\bY)$ denotes the $i$-th row of $\bY$, $\| \cdot \|$ denotes Euclidean distance, and the denominator $d_i+1$ is such that the SDS of a singleton node is well-defined. 

\begin{algorithm}[t!]
	\caption{Multi-centrality graph PCA (MC-GPCA)}
	\label{algo_strucutural_PCA}
	\begin{algorithmic}
		\State \textbf{Input:}  A graph $G=(\cV,\cE)$, desired dimension $q$
		\State \textbf{Output:} $n$ structural coordinates $\bY$ for each node in $G$
		\State 1. Extract $p$ structural vectors $\bX$ from $G$
		\State 2. Normalize each column of $\bX$ to have unit norm
		\State 3. Subtract row-wise empirical average from $\bX$
		\State 4. Solve the right singular vectors $\bV_q$ of $\bX$ 
		\State 5. $\bY=\bX \bV_q$
	\end{algorithmic}
\end{algorithm}

\begin{algorithm}[t!]
	\caption{Multi-centrality graph dictionary learning (MC-GDL)}
	\label{algo_graph_dictionary}
	\begin{algorithmic}
		\State \textbf{Input:}  A set of graphs $\{G_\ell \}_{\ell=1}^{g}$, number of atoms $K$, sparsity constraint $S$, number of highest SDS feature z
		\State \textbf{Output:}  graph structure dictionary $\bD$, coefficient matrix $\bC$
		\State 1. Obtain z highest SDS from (\ref{eqn_SDS}) for each graph as columns of $\bZ$
		\State 2. Subtract column-wise empirical average from $\bZ$
		\State 3. Perform K-SVD on $\bZ$ to obtain $\bD$ and  $\bC$		
	\end{algorithmic}
\end{algorithm}

\subsection{Multi-centrality graph dictionary learning (MC-GDL)}
Consider the case where a set of graphs $\{G_\ell \}_{\ell=1}^{g}$ is available, each possibly being of different graph size and connectivity pattern,
e.g., data from a cyber network at different time instances.
Multiple-centrality graph dictionary learning (MC-GDL) is proposed to learn a sparse structure representation of  $\{G_\ell \}_{\ell=1}^{g}$ by finding a dictionary $\bD$ consisting of $K$ atoms (columns of $\bD$) and an associated sparse coefficient matrix $\bC \in \mathbb{R}^{K \times g}$ such that the representation error $\|\bZ-\bD\bC\|_F$ is minimized while satisfying the column-wise sparsity constraints on $\bC$ that the number of nonzero entries of each column can not exceed a specified value $S$, where the columns in $\bZ$ are structural features of  $\{G_\ell \}_{\ell=1}^{g}$  and $\| \cdot\|_F$ denotes the Frobenious norm. 
Many different methods exist for solving the dictionary learning problem of estimating $\bD$ and $\bC$, often called the sparse coding problem \cite{lee2006efficient,jenatton2010proximal}. In this paper, we focus on a spectral method (K-SVD) of dictionary learning introduced in \cite{aharon2006KSVD}. The proposed MC-GDL selects 
the $z$ highest SDS from each graph as one column of $\bZ$ and applies K-SVD to find the dictionary and the corresponding coefficient matrix. The algorithm is  summarized in \textbf{Algorithm} \ref{algo_graph_dictionary}.

\section{Experiments and Cyber Intrusion Detection}
\label{sec_experiment_cyber}

\subsection{Illustration of sensitivity to structural changes on graphs}
Here we consider four similar graphs with different structural characteristics as displayed in Fig. \ref{Fig_SGPCA_demo} (a). From top to bottom, these four graphs represent high structural symmetry, reduced structural symmetry due to edge removal,  increase of the weight of edge (3,4), and change in edge direction. The extracted multi-centrality features are 1) graph walk statistics from 1 to 4 hops, and 2) the graph distance to node 1 (the reference node). It can be observed from Fig. \ref{Fig_SGPCA_demo} (b) that MC-GPCA can reflect structural perturbations, and total data variance is explained by one or two principal components. Moreover, the first principal component is shown to completer describe the network flow pattern for the directed example graph. Fig. \ref{Fig_SGPCA_demo} (c) shows that the graph distance feature adds discrimination power as the MC-GPCA scores are better differentiated.

%

\begin{table}[t]
	\centering
	\caption{Description of the University of New Brunswick (UNB) Intrusion Detection Evaluation Dataset \cite{shiravi2012toward}}
	\label{table_UNB_description}
	\begin{tabular}{c|c|c|c}
		\hline
		Dataset & \# nodes & \# edges & Description                                                                                  \\ \hline
		Day 1   & 5357     & 12887           & Normal activity                                                                              \\
		Day 2   & 2631     & 5614            & Normal activity                                                                              \\
		Day 3   & 3052     & 5406            & \begin{tabular}[c]{@{}c@{}}Infiltrating attack and \\ normal activity\end{tabular}           \\
		Day 4   & 8221     & 12594           & \begin{tabular}[c]{@{}c@{}}HTTP denial of service \\ attack and normal activity\end{tabular} \\
		Day 5   & 24062    & 32848           & \begin{tabular}[c]{@{}c@{}}Distributed denial of  \\ service attack using Botnet\end{tabular} \\
		Day 6   & 5638     & 13958           & Normal activity                                                                              \\
		Day 7   & 4738     & 11492           & \begin{tabular}[c]{@{}c@{}}Brute force SSH attack \\ and normal activity\end{tabular}        \\ \hline
	\end{tabular}
\end{table}

	\begin{figure}[t]
		\centering
		\begin{subfigure}[b]{0.23\textwidth}
			\includegraphics[width=\textwidth]{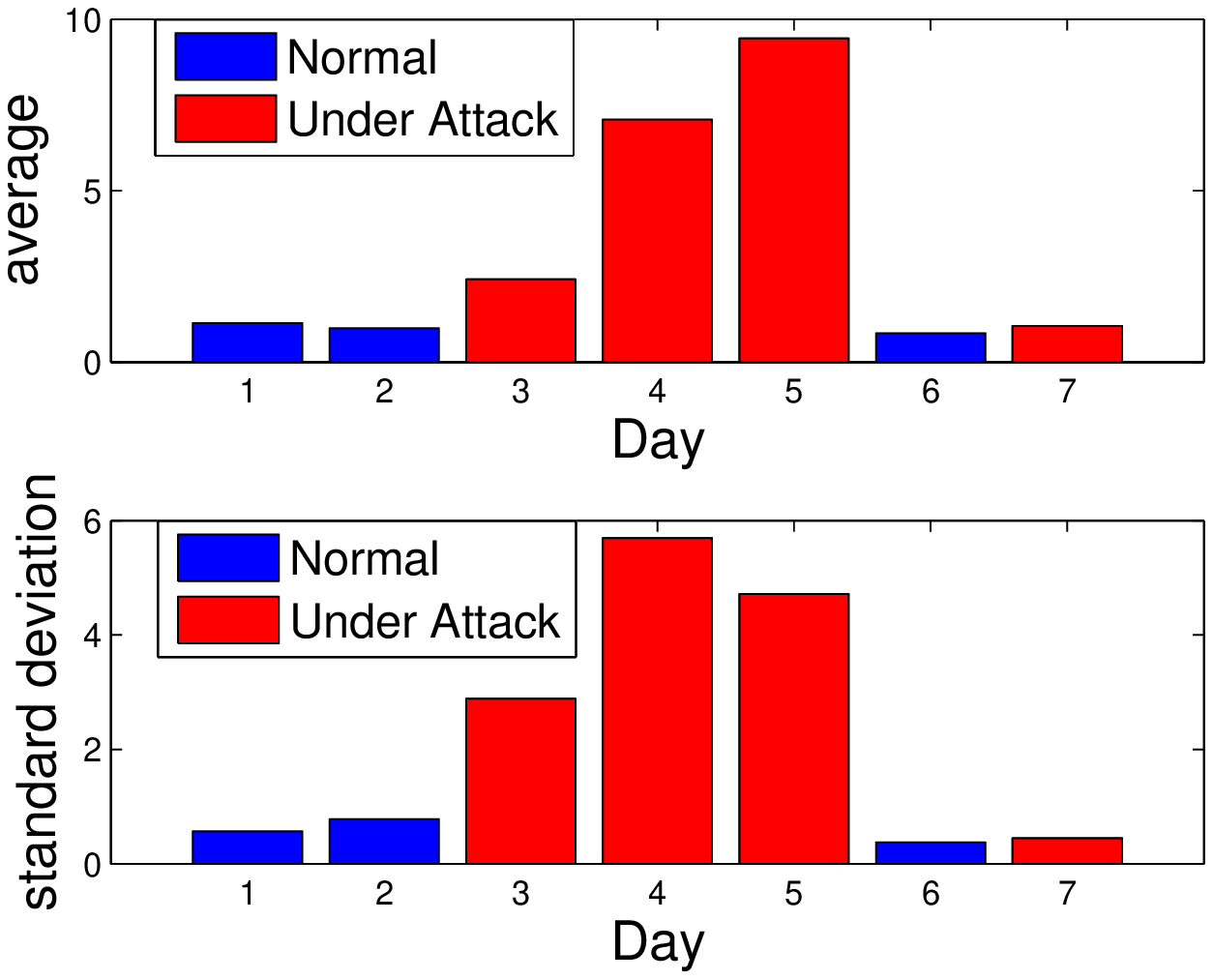}
			\caption{Graph-wise SDS statistic}
			\label{Fig_UNB_SGPCA}
		\end{subfigure}%
				\hspace{0.001cm}
		\centering
		\begin{subfigure}[b]{0.23\textwidth}
			\includegraphics[width=\textwidth]{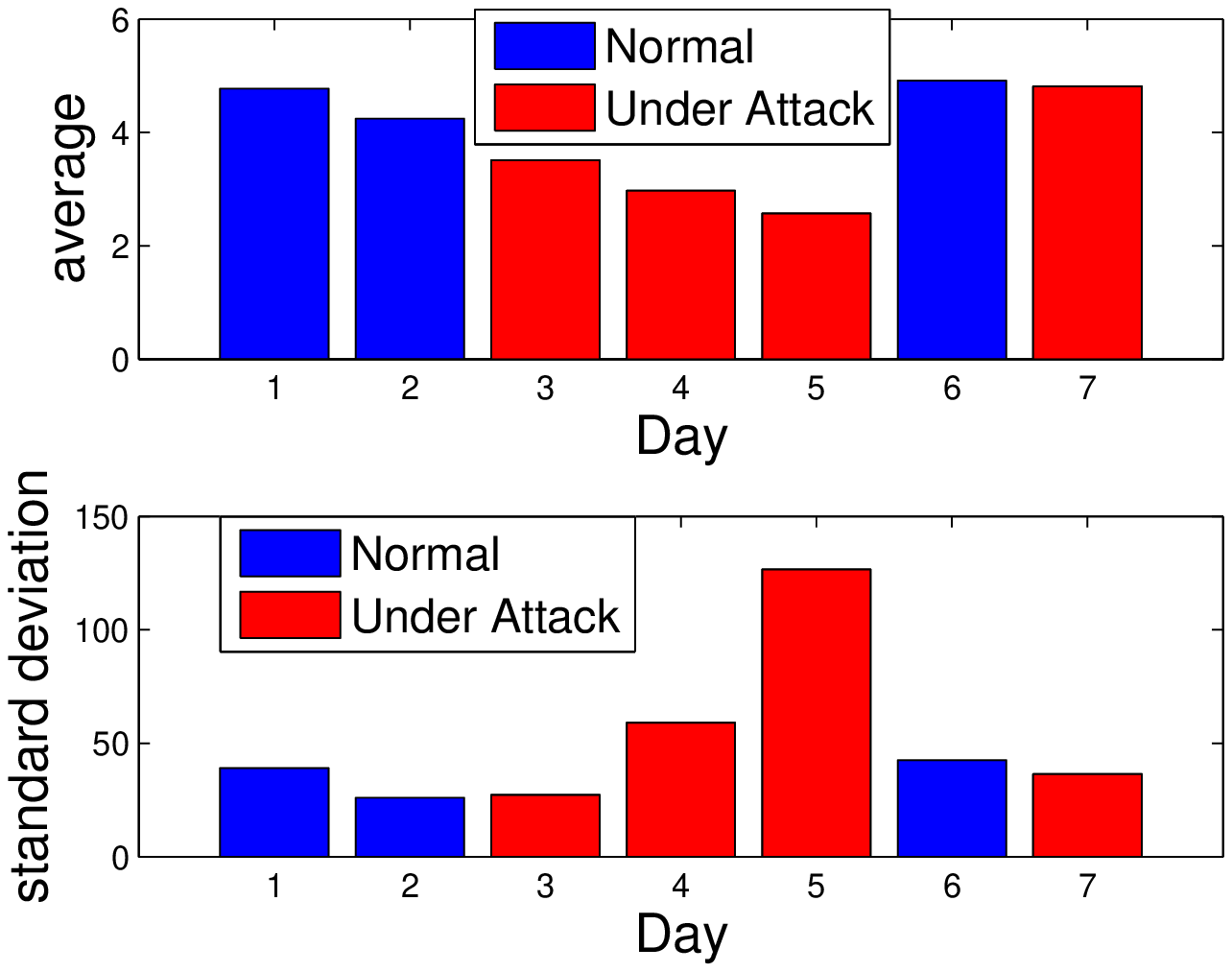}
			\caption{Graph-wise degree statistic}
			\label{Fig_UNB_Degree}
		\end{subfigure}%
			\hspace{0.001cm}
					\begin{subfigure}[b]{0.23\textwidth}
						\includegraphics[width=\textwidth]{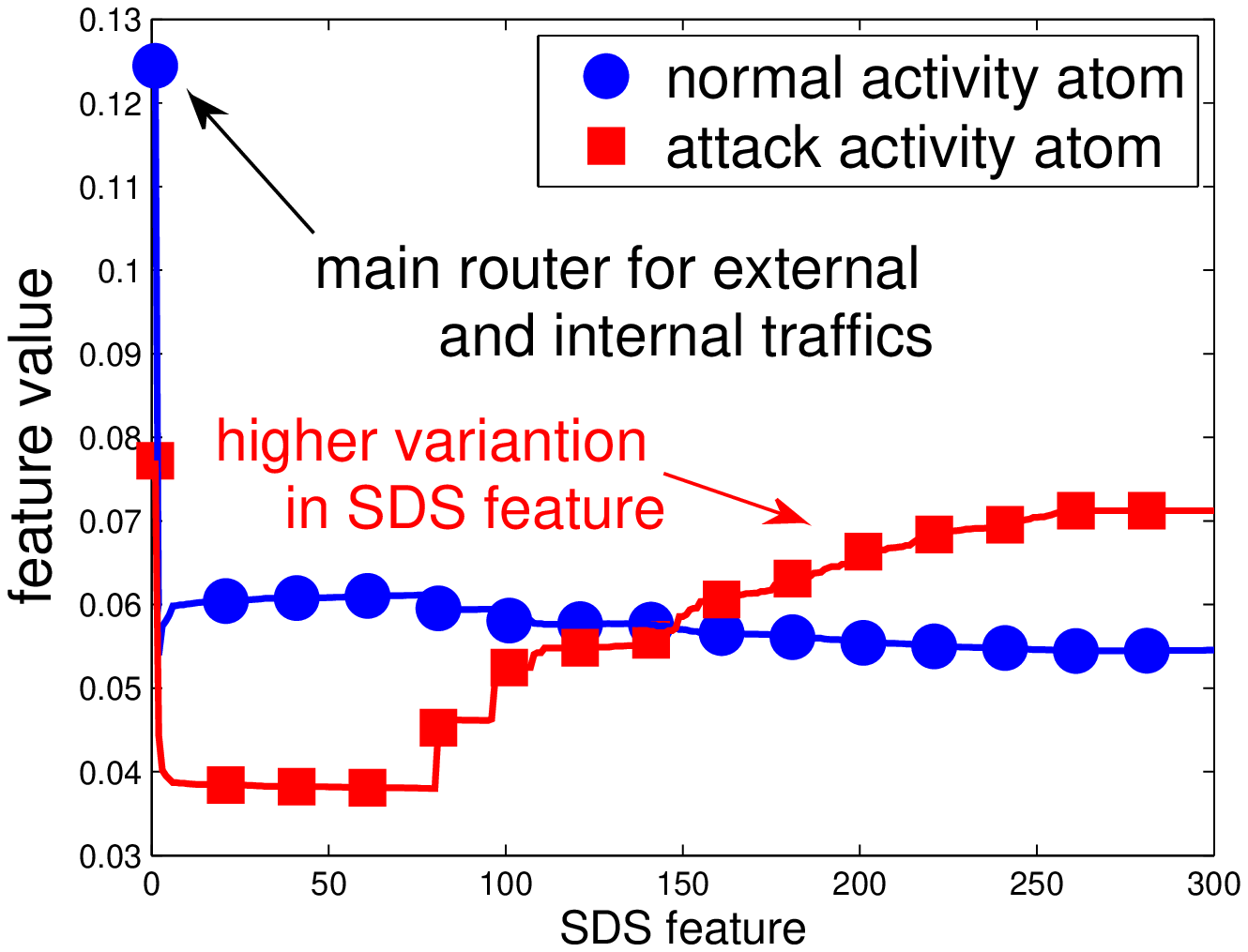}
						\caption{Dictionary from MC-GDL}
					\end{subfigure}%
							\hspace{0.001cm}
					\centering
					\begin{subfigure}[b]{0.23\textwidth}
						\includegraphics[width=\textwidth]{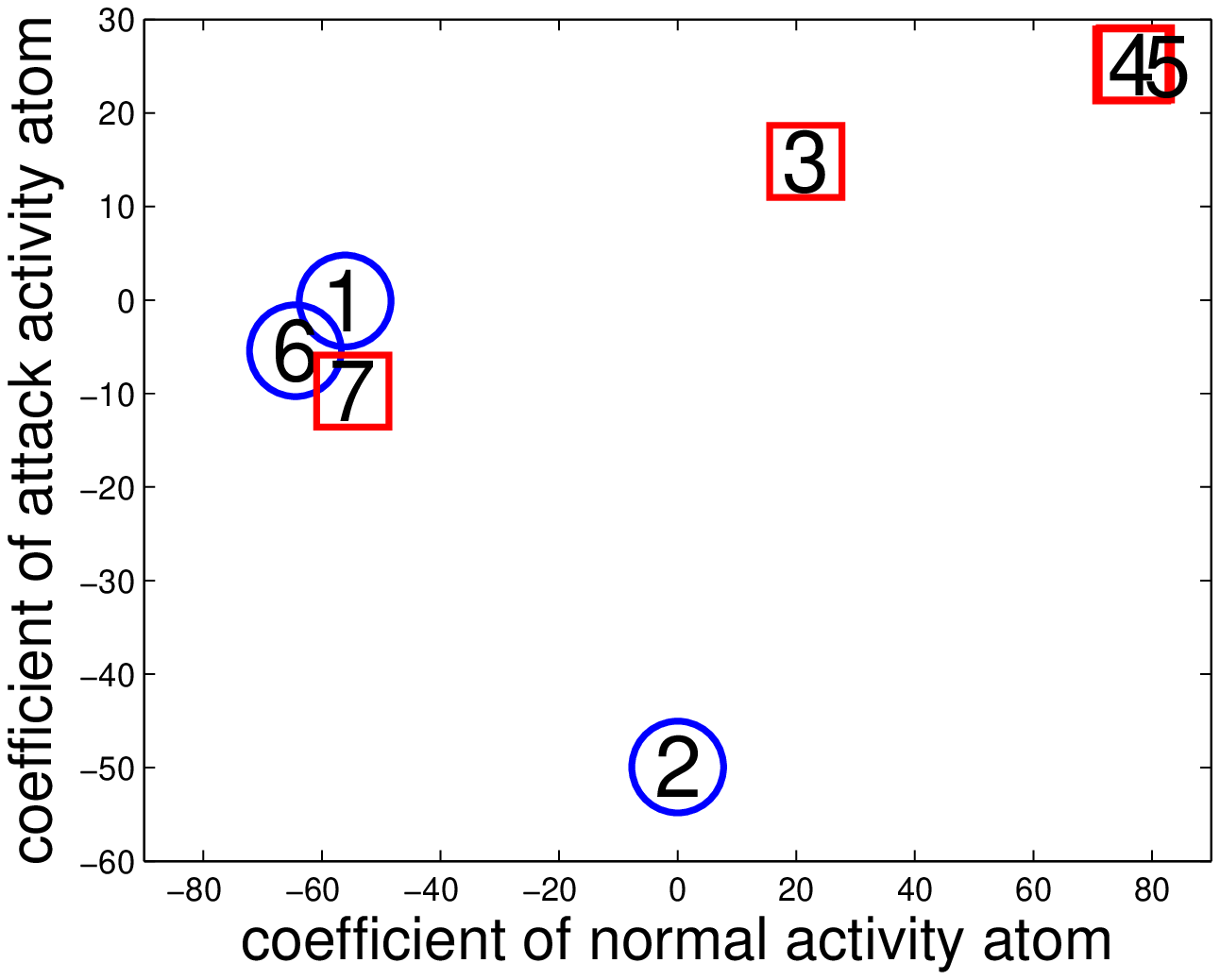}
						\caption{Coefficients from MC-GDL}
						\label{Fig_UNB_Dict_Coeff}
					\end{subfigure}%
		\caption{Cyber intrusion detection on the UNB dataset. MC-GPCA and MC-GDL are shown to be effective indicators of cyber attacks.		
		}
		\label{Fig_UNB}
							\vspace{-6mm}
	\end{figure}


\subsection{Cyber intrusion detection}
The UNB intrusion detection evaluation dataset \cite{shiravi2012toward} described in Table \ref{table_UNB_description} is a collection of directed cyber network graphs where each node is a host (machine) in a cyber system and an edge indicates the existence of communication between hosts. No information beyond graph topology is used for analysis.
The extracted multi-centrality features are 1) graph walk statistics from 1 to 20 hops, 2) all centrality measures introduced in Sec. \ref{subsec_centrality_measure} (edge directions are omitted for computing LFVC), and 3) graph distances to 10 reference nodes of highest degree, resulting in $p=56$ features (columns of $\bX$).
Fig. \ref{Fig_UNB} (a) shows that the proposed SDS statistic (Eqn. (\ref{eqn_SDS})) with $q=2$ principal components from MC-GPCA. The SDS statistics  are similar over days without attacks, whereas they are significantly higher in days under attacks that induce anomalous connectivity patterns (i.e. Days 3, 4 and 5). On the other hand degree statistic (Fig. \ref{Fig_UNB} (b)) fails to be a valid indicator of cyber attacks. 
	The SDS statistic fails to detect the SSH attack (Day 7) since it is a password attack that takes place only between a single host and a single server.

We applied MC-GDL to the entire UNB database of graphs to learn a dictionary that spans the dataset. For this implementation of MC-GDL 
we select $K=2$ atoms, $z=300$ SDS features and $S=2$ sparsity level.  The two learned structural atoms in 
	Fig. \ref{Fig_UNB} (c) can be interpreted as a normal activity atom consisting of identical SDS features except for one spike accounting for the main router and an attack activity atom of higher variance in SDS features. 	
	The corresponding coefficients in Fig. \ref{Fig_UNB} (d) reflect the mixture portion of these atoms and they can be used for attack classification. For instance, $K$-means clustering with $2$ clusters identifies Days 3, 4 and 5 as being anomalous  and thus under attack.

\section{Conclusion}
This paper proposes PCA and dictionary learning graph decomposition methods that are based on multi-centrality features of the graph. 
The proposed methods can reflect structural perturbations in graph symmetry, edge weight and edge direction. When applied to cyber intrusion detection, our experiments show 
that MC-GPCA and MC-GDL can effectively detect attacks on the network.

\clearpage
\bibliographystyle{IEEEtran}
\bibliography{IEEEabrv,CPY_ref_20150925}

\end{document}